\documentclass[prb,aps,twocolumn,dcolumn,superscriptaddress,showpacs,amsmath,amssymb,showkeys]{revtex4}

\usepackage{graphicx}
\usepackage{graphicx}
\begin{document}
\title{How systems of single-molecule magnets magnetize at low temperatures}
\author{Julio F. Fern\'andez}
\affiliation{ICMA, CSIC and Universidad de Zaragoza, 50009-Zaragoza, Spain}
\email[E-mail address: ] {JFF@Pipe.Unizar.Es}
\homepage[ URL: ] {http://Pipe.Unizar.Es/~jff}
\author{Juan J. Alonso}
\affiliation{F\'{\i}sica Aplicada I, Universidad de M\'alaga,
29071-M\'alaga, Spain}
\email[E-mail address: ] {jjalonso@Darnitsa.Cie.Uma.Es}
\date{\today}

\begin{abstract}     
We model magnetization processes that take place through tunneling in 
crystals of single--molecule magnets, such as Mn$_{12}$
and Fe$_8$. These processes take place when a field $H$ is applied 
after quenching to very low temperatures. Magnetic dipolar
interactions and spin flipping rules are essential ingredients of the 
model. The results obtained follow
from Monte Carlo simulations and from the stochastic model we propose 
for dipole field diffusion. Correlations
established before quenching are shown to later drive the 
magnetization process. We also show that in simple cubic lattices,
$m\propto \sqrt{t}$ at time $t$ after $H$ is applied, as observed in 
Fe$_{8}$, but only for $1+2\log _{10}(h_d/h_w)$ time decades,
where $h_d$ is some near--neighbor magnetic dipolar field and a spin 
reversal can occur only if the magnetic field acting on it is
within some field window ($-h_w,h_w$). However, the $\sqrt{t}$ 
behavior is not universal. For BCC and FCC lattices, $m\propto t^p$,
but $p\simeq 0.7$. An expression for $p$ in terms of lattice 
parameters is derived.
At later times the magnetization levels off to a constant value. All 
these processes take place at approximately constant magnetic
energy if the annealing energy
$\varepsilon_a$ is larger than the tunneling window's energy width 
(i.e., if $\varepsilon_a\gtrsim g\mu_Bh_wS$). Thermal processes
come in only later on to drive further magnetization growth.

\end{abstract}
\maketitle
\section{introduction}
Some magnetic clusters, such as Fe$_8$ and Mn$_{12}$, that make up 
the core of some organometallic molecules, behave at low
temperatures as large single spins. Accordingly, these molecules have 
come to be known as single--molecule magnets (SMM's). \cite{SD}
In crystals, magnetic anisotropy barriers, of energy $U$, inhibit 
magnetic relaxation of SMM's, which can consequently proceed only by
tunneling under the barriers. Magnetic quantum tunneling (MQT) was 
first observed to take place through thermally excited
states,\cite{friedm} but temperature--independent ``pure'' MQT was 
observed shortly thereafter.\cite{sangregorio} Dipolar interactions
then play an essential role. They can give rise, upon tunneling, to 
Zeeman energy changes of nearly 1 K. This exceeds by many orders
of magnitude the ground state tunnel splitting energies
$\Delta$ that would follow for Fe$_8$ and Mn$_{12}$ from 
perturbations by anisotropies.\cite{highan} Energy conservation
would make pure MQT impossible for the vast majority of spins in the 
system. Hyperfine interactions between the tunneling electronic
spins of interest and nuclear spins open up a fairly large tunneling 
window of energy
$\varepsilon_w$ such that tunneling can occur if the Zeeman energy 
change $2\varepsilon_h$ upon tunneling is not much larger than
$\varepsilon_w$.\cite{PS} More precisely, the tunneling rate
$\Gamma^\prime$ for spins at very low temperature is given by
\begin{equation}
\Gamma^\prime (\varepsilon_h) \simeq
\Gamma\;\eta (\varepsilon_h/\varepsilon_w),
\label{gamma}
\end{equation}
where $\Gamma$ is some rate (whose value is not important for our 
purposes), $\eta (x)\sim 1$ for $\mid x \mid < 1$, $\eta
(x)\sim 0$ for $x> 1$, and $\varepsilon_w\gg\Delta$. Other theories 
for MQT of SMM at very low temperatures have also been
proposed.
\cite{chudno} We adopt Eq. (\ref{gamma}) here, regardless of theory 
or physical mechanism behind it. We let $\eta
(x)= 1$ for $\mid x\mid < 1$ and $\eta (x)=0$ for $x\ge 1$, and refer 
to $\varepsilon_w$ as the tunnel energy window.

The interesting early time relaxation, $m(0)-m(t)\propto \sqrt {t}$, 
of the magnetization $m$ of a system of SMM's that is fully
polarized initially has been predicted,\cite{PS}
observed experimentally,\cite{exp,expb} further explained,\cite{vill} 
and widely discussed.\cite{comm} An unpredicted related
phenomenon was later observed by Wernsdorfer et al.:\cite{ww1} the 
magnetization $m$ of a system of Fe$_8$ SMM's increases
as $\sqrt{t}$, where $t$ is the time after a weak magnetic field is 
applied to an initially unpolarized system. There are important
differences between the early time relaxation of the magnetization of 
a system that is fully polarized initially\cite{PS,exp} and the
magnetization process from an initially unpolarized state.\cite{ww1} 
Whereas the former effect depends
crucially on system shape, the latter does not. Interesting questions
arise: is this a universal effect to be found in all MQT experiments? 
If not, what does it depend on? How many time
decades does the $\sqrt{t}$ regime cover? What is the final steady 
state magnetization? We have reported in Ref. \onlinecite{Letter}
a few results from MC simulations that answer some of these 
questions. The following problems were however not addressed in Ref.
\onlinecite{Letter}: (a) the crucial effect that energy transfer (or 
the lack of it) between the magnetic system and the
lattice has on the nature of the magnetization process; (b) a closely 
related problem, how the
magnetization relaxes in zero field from an initially weakly 
polarized state. In addition, some results from our theory were given 
in
Ref. \onlinecite{Letter}, but the theory itself was not.

In this paper, we report extensive results from Monte Carlo (MC) 
simulations, an approximate theory, and results that follow from it.
The notation is first specified. Unless otherwise stated, all 
energies and magnetic fields are given in terms of $\varepsilon_d$ and
$h_d$, respectively, where $h_d=g\mu_BS/a^3$, $a$ is the cubic 
lattice parameter, $g$ is the gyromagnetic ratio, $\mu_B$
is the Bohr magneton, $S$ is the spin size, and 
$\varepsilon_d=h_dg\mu_BS$. Temperatures are given in units of
$\varepsilon_d/k_B$. For comparison purposes, the ordering 
temperature $T_0$ and the ground state energy $\varepsilon_0$ are 
given in
Table I for an Ising model with dipolar interactions in simple cubic 
(SC), body--centered--cubic (BCC), and face--centered--cubic
(FCC) lattices. We let $\langle h^2 \rangle _0$ stand for the 
mean--squared value of the magnetic dipolar field $h$ for random spin
configurations, and $\sigma$ for $1/[\sqrt{2\pi}p(0)]$, where $p(0)$ 
is the probability density function (PDF) that the field on a
randomly chosen site is $0$ when the spin configuration is random.\cite{gauss0}

\begin{figure}
\includegraphics[width=8cm]{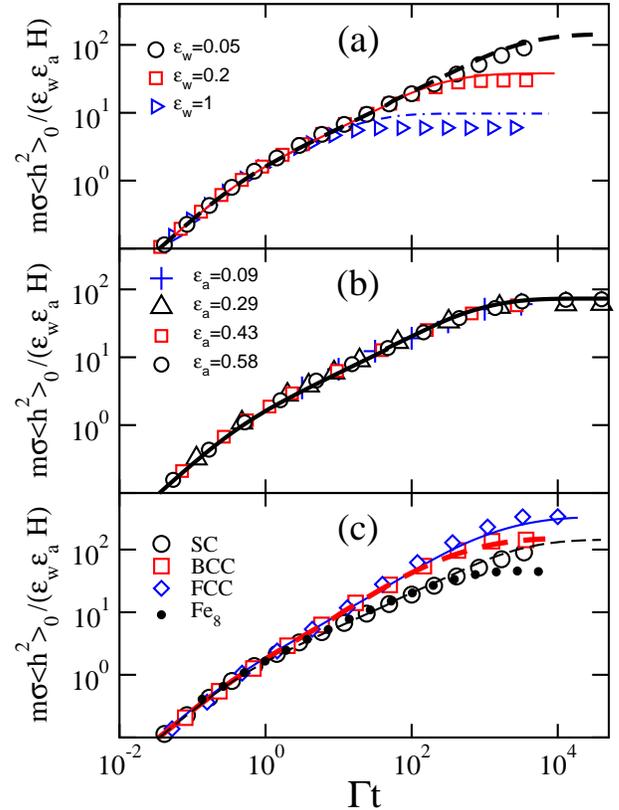}
\caption{(a) $m\sigma\langle 
h^2\rangle_0/(\varepsilon_w\varepsilon_aH)$ versus $\Gamma t$ for an
applied external field $H=1$ and the shown values of $\varepsilon_w$. 
Symbols stand for Monte Carlo results, but dashed, full, and
dot--dashed lines stand for results obtained from Eq. (\ref{mfinal2}) for
$\varepsilon_w=0.05, 0.2$, and $1$, respectively.
Before field $H$ was applied, thermalization took place from an 
initially disordered configuration up to the time when the
energy $-\varepsilon_a$ reached the value
$-0.58$. Data points for $\Gamma t < 4$ follow from $4\times 10^4$ 
runs for systems of $16\times 16\times 16$ spins on a SC lattice.
For
$\Gamma t > 4$, all data points are for averages over $4\times 10^5$ 
runs of systems of $8\times 8\times 8$ spins in a SC lattice.
(b) Same as in (a) but for the shown values of $\varepsilon_a$ and 
$\varepsilon_w=0.1$. The continuous line is from the numerical
solution of Eq. (\ref{mfinal2}). (c) Same as in (a) and (b), but for 
different lattices. Symbols stand for MC results.
Lines stand for results from theory: dashed is for SC and Fe$_8$ 
lattices, dot--dashed and full lines are for BCC and FCC lattices
respectively. For SC, BCC, FCC, and Fe$_8$ lattices, partial 
thermalization previously took place at $T_a=10$,
$20$, $60$, and $1.06$K, till $\varepsilon_a=0.31$, $0.28$, $0.53$, 
and $0.15$K, respectively. (For easy comparison with table
I, units for Fe$_8$ are given in Kelvin --but not for cubic 
lattices.) We used $\varepsilon_w=0.05$ for all cubic lattices and
$\varepsilon_w=11$mK for Fe$_8$.}
\label{funo}
\end{figure}

The main results obtained follow. We show that the nonlinear in time 
magnetization arises from
correlations that develop between spins and local magnetic dipolar 
fields, while cooling to low temperatures, before finally
quenching to experiment. Furthermore, only the final energy 
$-\varepsilon_a$ reached
just before quenching matters about the cooling protocol. More 
specifically, after quenching and applying a field $H\lesssim1$ at
$t=0$,
\begin{equation}
m(t)\simeq  b\frac{\varepsilon_a \varepsilon_w H}{\langle 
h^2\rangle_0\sigma}F(\Gamma t,\sigma/\varepsilon_w,\sigma /h_0),
\label{mdet}
\end{equation}
where $b\simeq 4\sqrt{2/\pi}$,
\begin{eqnarray}
F&\simeq & \Gamma t \;\;\text{ \hspace{1.2cm}  for } \Gamma t\lesssim 1\\
\label{Fdeta}
F&\simeq & 0.7(\Gamma t)^p \text{ \hspace{0.5cm}  for } 
1\lesssim\Gamma t\lesssim (\sigma/\epsilon_w)^{1/p}\\
F &\simeq &\frac{1}{2}\sqrt{\frac{\pi}{2}}\,\sigma\varepsilon_w^{-1} 
\text{ \hspace{0.5cm}  for }
(\sigma/\varepsilon_w)^{1/p}\lesssim
\Gamma t,
\label{Fdetb}
\end{eqnarray}
$h_0=(8\pi ^2/3^{5/2})g\mu_BS\rho_v$, $\rho _v$ is the number of spin 
sites per unit volume, $p\simeq 0.5$
for  SC lattices, and
$p\simeq 0.7$ for BCC and FCC lattices. These results, shown 
graphically in Fig. \ref{funo} are obtained from MC simulations in 
which
$\varepsilon_a\gtrsim \varepsilon_w$ as well as from the theory given 
below. The theory also gives
\begin{equation}
\frac{\sin \pi p}{p}=\sqrt{2\pi}\frac{\sigma}{h_0}.
\label{1}
\end{equation}
Additional results are mentioned in the plan of the paper below.

The plan of the paper is as follows. The model and Monte Carlo 
simulations are described in Sect. \ref{mods}. We also explain in
Sect. \ref{mods} how we simulate constant energy processes. Section 
\ref{theor} is
devoted to the theory. As an introduction to the rest of the section, 
qualitative arguments are given in Sec. \ref{over}. The
difference between the probability density functions for up--spins 
and down--spins with a dipolar field $h$ acting on them, which
develop during the stage before quenching to very low temperatures, 
is derived in Sec. \ref{anneal}. The theory for the
magnetization process that takes place after a magnetic field $H$ is 
applied, soon after quenching, is given in Sec. \ref{mproc} and
the Appendix. Field functions that are defined below develop in time 
holes that have been observed experimentally\cite{ww1} and are
intimately related to the magnetization process. How these holes 
develop in time is the subject of Sec. \ref{TWI}. In Sec.
\ref{depol}, we study the relaxation in zero field of the 
magnetization of systems that have been previously cooled in weak 
fields. In
Sec. \ref{heat} we show how the magnetization crosses over to a 
linear in time behavior long after a field is applied if energy
exchange between the magnetic system and heat reservoir takes place. 
Finally, the results obtained are discussed in Sec.
\ref{dis}.

\begin{table}
\caption{\label{tabla}$\sqrt{\langle h^2\rangle_0}$ stands for the 
rms spatial average of the dipolar field, $\sigma\equiv
[\sqrt{2\pi}p(0)]^{-1}$, and
$p(0)$ stands for the PDF at $h=0$ on a randomly chosen site, both 
for completely random
spin configurations. Except for Fe$_8$, they are both in units of
$g\mu_BSa^{-3}$, and follow from present MC simualtions; $a$ is the 
lattice constant for cubic lattices.
$T_0$ and $\varepsilon_0$ stand for the ordering temperature and 
ground state energy,
respectively. Their values are taken from the shown references.
$k_BT$ and $\varepsilon_0$ are, except for Fe$_8$, in units of
$(g\mu_BS)^2 a^{-3}$.}
\begin{ruledtabular}
\begin{tabular}{|c|l|l|l|c|l|}
LATTICE        &$n_b$          & $\sqrt{\langle h^2\rangle_0} $ 
&$\sigma$  &$T_0$ & $\varepsilon_0$\\ \colrule
SC\footnotemark[1] & 1 &3.655        & 3.83(2)   & 2.50(5) & -2.68(1)      \\
BCC\footnotemark[1] & 2 & 3.864      & 4.03(2)   & 5.8(2)  & -4.0(1)       \\
FCC\footnotemark[1] & 4 & 8.303      & 8.44(2)   & 11.3(3) & -7.5(1)       \\
Fe$_8$\footnotemark[2] & 1 & 46(1) mT & 31(1) mT & 0.4(1) K & -0.51(2) K\\
\end{tabular}
\end{ruledtabular}
\footnotemark[1] These numbers for $T_0$ and $\varepsilon_0$ are from 
Ref. \onlinecite{dipole}.\\
\footnotemark[2] We assume an easy anisotropy axis as given in 
\onlinecite{JF,barra}. $T_0$ and $\varepsilon_0$ are from Ref.
\onlinecite{erratum}.
\end{table}

\section{Model and simulation}
\label{mods}
We use the MC method to simulate magnetic relaxation of Ising systems 
of $\pm S$ spins, on simple cubic lattices with periodic
boundary conditions (PBC), that interact through magnetic dipolar 
fields and flip under rules to be specified below.\cite{Ising}
In our PBC scheme, two spins interact whether they are in the same 
box or in different replicated boxes, but
a spin at x,y,z interacts with a spin at x',y',z' only if
$-L_x/2\leq x-x^\prime <L_x/2$, $-L_y/2\leq y-y^\prime <L_y/2$, and 
$-L_z/2\leq z-z^\prime <L_z/2$, where $L_x$,
$L_y$, and $L_z$ are the sides of the box--like systems we simulate. 
Thus each spin interacts with $N-1$ spins, where $N$ is the
number of spins in the system. This scheme has been tested against a 
free--boundary scheme,\cite{dipole,Alonso} in which all spins in
the system are allowed to interact, and found satisfactory when the 
system is only weakly polarized (or not at all), as is expected
from Griffith's theorem.\cite{griff}

The system is first allowed to evolve towards thermal equilibrium at 
some ``high'' temperature $T_a$. We assume
$k_BT_a\gtrsim U/10$, which implies that spin reversals then take 
place mostly through overbarrier processes.
Accordingly, spin
flips are then governed by detailed balance rules, and Eq. 
(\ref{gamma}) is not enforced. For reasons that become clear below, 
we also
impose the restriction
$T_a\gtrsim T_0$, where $T_0$ is the long-range ordering temperature. 
One may think of this first process as a waiting stage that the
systems may have to undergo in the cooling process, before quenching 
to a lower temperature where a tunneling experiment (as in
Ref.~\onlinecite{ww1}) can be later performed.

\begin{figure}
\includegraphics[width=8cm]{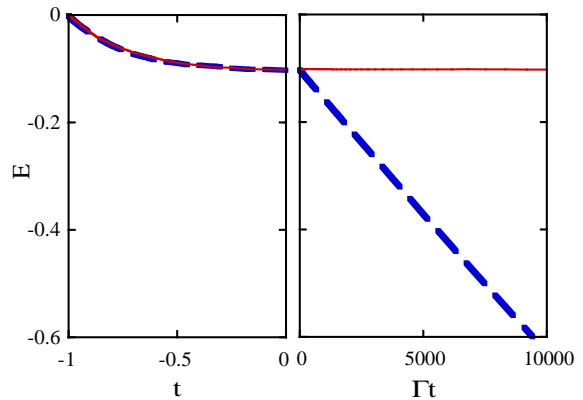}
\caption{Energy versus time in the annealing stage, when $t<0$, and 
after quenching, when $t>0$. The time scale in the annealing
stage, when $t<0$, is such that the rate for energy lowering spin 
flips is $1$. In the annealing stage $T=62$ in
both cases. For comparison purposes, the ordering temperature $T_0$ 
and the ground state energy $\varepsilon_0$ are given in Table I.
For $t>0$, the continuous and dashed lines are for constant energy 
and for an isothermal process at $T=1$, respectively.
Equation (1) was enforced for both evolutions, but, as everywhere 
else, only when $t>0$. The constant energy process was simulated
assuming detailed balance with $T_u=63$. These data are for systems 
$16\times 16\times 16$ spins, with $\varepsilon_w=0.1$, averaged
over $10^3$ runs.}
\label{evst}
\end{figure}

Let the time when this first stage is ended by quenching to a 
temperature below roughly
$0.1U/(Sk_B)$ be $t=0$.\cite{sangregorio,ree} Accordingly, Eq. (\ref{gamma}) is
then enforced on all spin flips for $t>0$. As for detailed balance, 
we then proceed as follows. We may assume (1) that thermalization
of a SMM system with the lattice does not take place (i.e., the 
energy is constant) at very low temperatures, or (2) that heat is
readily exchanged with the lattice, as is sometimes the 
case\cite{metes}. In the latter case, we follow the standard
procedure. Monte Carlo results illustrate these two cases in Fig. 
\ref{evst}. We fulfill the constant energy condition by enforcing
detailed balance but using an appropriately chosen pseudotemperature 
$T_u$. [From an expression below Eq. (\ref{x2}),
$k_BT_u\approx \langle h^2 \rangle_0 /2\varepsilon_a$.\cite{xxx} Note 
that $T_u\geq T_a$, since $-\varepsilon_a$ cannot be smaller
than the equilibrium energy at $T_a$]. We have checked that the mean 
energy is indeed constant under this rule, as illustrated in Fig.
\ref{evst}. Unless otherwise stated (as in sect. \ref{heat}), results 
reported below are for constant energy processes.

MC results for the time evolution of $m\langle h^2 
\rangle_0\sigma/(\varepsilon_w\varepsilon_aH)$, after a field $H=1$ 
is applied upon
quenching, are shown in Fig. \ref{funo}(a) for various values of 
$\varepsilon_w$. Before quenching, the system was thermalized at
$T_a=10/k_B$ for some time till the energy per spin reached the value 
$-0.58$. Clearly, $m$ scales with $\varepsilon_w$ up to a
crossover time of, roughly,
$\Gamma ^{-1}\sigma^2/\varepsilon_w^2$, where $m$ levels off. Within 
the time range $1\lesssim \Gamma t \lesssim
\sigma^2/\varepsilon_w^2$, $m\propto \sqrt{t}$. Monte Carlo results 
that show how $m$ scales with $\varepsilon_a$ are exhibited in
Fig. \ref{funo}(b) for $\varepsilon_w=0.1$. Scaling holds also for 
the leveling off value $m_e$ of $m$. Equations (2-5) are
inferred from these graphs, as well as from the theory given below, 
from which some of the constants are obtained.

\section{theory}
\label{theor}
In this section, we try to understand the results of Sect. II and go 
somewhat beyond.

\subsection{Overview}
\label{over}
\begin{figure}
\includegraphics[width=8cm]{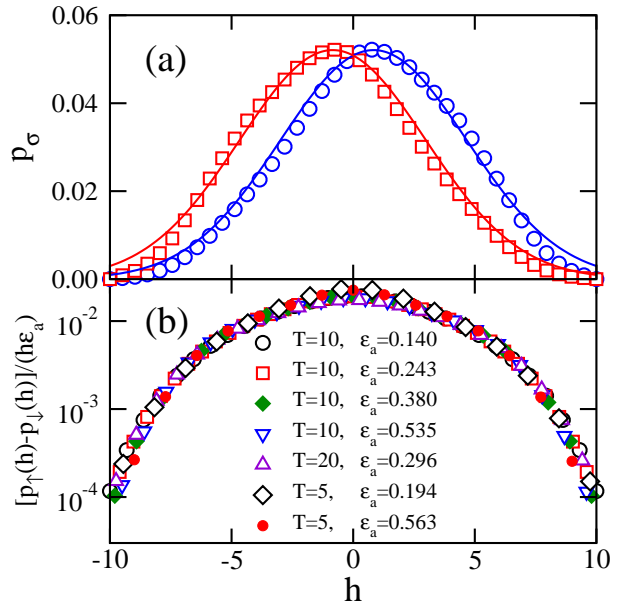}
\caption{(a) $p_\sigma (h)$, versus $h$. All data points are for a 
system of $16\times 16\times 16$ spins. $\Box$ and $\circ$ stand
for down--spin and up--spin, respectively. Full lines are for 
Gaussian functions centered on $h=0.84$ and $h=-0.84$, with standard
deviations of $3.83$. The system was cooled to a temperature 
$T_a=10$, from an infinite
temperature, and allowed to evolve towards equilibrium (but not to reach it) 
till its energy reached the value
$-\varepsilon_a=-0.42$. For comparison purposes see
$T_0$ and the ground state energy values in Table I, and the energy 
is given by $(1/2)\int h[p_\uparrow
(h)-p_\downarrow (h)] dh $ when no external field is applied. All 
points stand for averages over
$1.5\times 10^5$ runs. (b) $[p_\uparrow (h)-p_\downarrow 
(h)]/(h\varepsilon_a)$ versus $h$
for the shown values of $\varepsilon_a$.
No results are shown in the interval $-0.1<h  <0.1$, as division by
$h$ in the $h\sim 0$ neighborhood enhances errors greatly.}
\label{fdos}
\end{figure}

Rough arguments that can serve to introduce the derivations that 
follow in subsections below are given in this subsection. Before
quenching to very low temperatures, the system is for some time at 
some temperature $T_a$ that is above the ordering temperature
$T_0$. In addition, $k_BT_a\gtrsim U/10$. Overbarrier relaxation can 
then take place, and
$\tau =\tau_0 \exp (U/k_BT)$ follows from Arrehnius' law. 
Consequently, spin flipping readily takes place in the laboratory 
within a
second's time if $\tau_0\lesssim 10^{-5}$ s. Some correlation between 
spin $s_i$ and field $h_i$ at each site $i$ can therefore be
established, but no long-range order can develop as long as 
$T_a\gtrsim T_0$. Thus, immediately after quenching, the joint
probability density to find a spin up $p_\uparrow (h)$ with a field 
$h$ acting on it is larger than $p_\downarrow (h)$ if $h>0$ and
vice versa. This is illustrated in Fig. \ref{fdos}(a). As the system 
evolves towards equilibrium before quenching, we expect
$f(h)\equiv  p_\downarrow (h)-p_\uparrow (h)$ to increase. It is 
shown in the following section and illustrated in Fig.
\ref{fdos}(b) with MC results, that thermalization effects prior to 
cooling to very low temperatures turn out to be rather simple:
$f(h)\propto \varepsilon_ahp(h)$ if $\varepsilon_a\ll\mid 
\varepsilon_0\mid$, where $p(h)=p_\uparrow (h)+p_\downarrow (h)$ and
$\varepsilon_0$ is the ground--state energy
($-2.68$ for a SC lattice). Since $m=-\int dh
f(h)$ (from the definition of $f$), $m\propto \varepsilon_a$ is 
expected to ensue after quenching and application of an external
field.

\begin{figure}
\includegraphics[width=8cm]{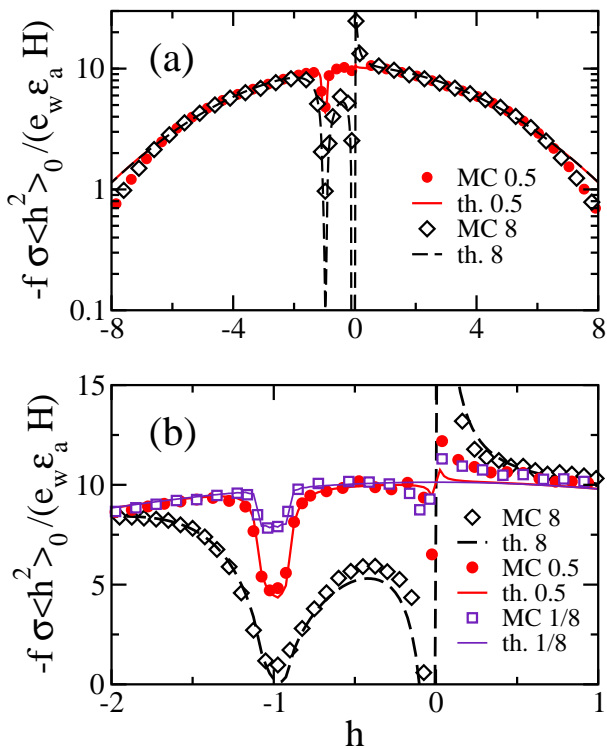}
\caption{ (a) $f\sigma\langle 
h^2\rangle_0/(\varepsilon_a\varepsilon_wh)$, at the shown times, for 
a system of $16\times 16\times 16$
spins, after it was thermalized at
$T=10$ till $\varepsilon_a\simeq 0.51$, then abruptly cooled, when 
$H=1$ was applied. Closed (open) symbols stand for MC (theory)
results. A tunnel window $\varepsilon_w=0.1$ was
enforced. All points stand for averages over $6\times 10^5$ 
histories. (b) Same as in (a) but for the shown values of time. All
results obtained from theory for
$f(h,t)$, that is, from Eq. (\ref{mfinal2}) were multiplied by $1.27$ 
in parts (a) and (b). The need for this correction quite
likely comes from second order effects in $\varepsilon_a/\sigma$ not 
included in the theory (see Sec. \ref{anneal}). }
\label{ftres}
\end{figure}

After quenching in the laboratory to a temperature below 
approximately $0.1U/S$, spin reversals take place only by tunneling
between $S_z=-S$ and $S_z=+S$ states. For this, a spin must be within 
the ``tunnel window'' (TW), defined by $\eta$ in Eq. (1).
Let some field $H$, fulfilling $H^2\ll\langle h^2\rangle_0 $, be 
applied after quenching. Then, $h+H$ must be within the TW,
that is, only spins with a magnetic dipolar field within a TW 
centered on $h=-H$ can flip. Consider some $H>0$. A look at Fig.
\ref{fdos}(a) shows that $p_\downarrow (-H)>p_\uparrow (-H)$ then, 
which implies that $m$ must increase with time after $t=0$, since
the number of upward flips minus the number of downward flips is 
proportional to $p_\downarrow (h)-p_\uparrow (h)$ within the TW (the
Zeeman energy for
$h=-H$ vanishes and we are assuming
$k_BT\gg\varepsilon_w$). We therefore expect $p_\downarrow 
(h)\rightarrow p_\uparrow (h)$ with time if $h$ is within the TW, but 
both
$p_\downarrow (h)$ and $p_\uparrow (h)$ to remain approximately 
constant for some time if $h$ is outside the TW. Monte Carlo
results illustrate this behavior in Figs. \ref{ftres}(a) and (b). 
This is much as in the experiments of Wernsdorfer et al.\cite{ww1}
In addition, inspection of Fig. \ref{fdos}(a) suggests that
$f(h)\propto h$ if $h^2 \ll\langle h^2\rangle_0 $, whence $m\propto 
H$ also follows. Finally,  since the number of spin flips
increases linearly with
$\varepsilon_w$ if $\varepsilon_w^2\ll \langle h^2\rangle_0$, we 
expect $m\propto \varepsilon_w$. This is all as in Eq. (2).

The field $h$ acting on a spin does not remain long within the TW 
after the spin flips. Changing fields produced by other spin
flips bring $h$ out of the TW, much as in a random walk. Thus, the 
``hole'' (or ``well'') exhibited in Fig. \ref{ftres}(a) widens as
in a diffusion process. This is treated in detail in Sec. \ref{TWI} 
and illustrated in Fig. \ref{ftres}(b). Initially, we expect
the well's width
$\delta w$ to increase linearly with time, since a dilute, spatially 
random distribution of spins gives a Lorentzian field
distribution,\cite{PWA} with a width that is proportional to the 
concentration. However, as the concentration $c$ of flipped spins
grows, the corresponding field distribution becomes Gaussian, with 
$\delta w\propto \sqrt{c}$. Consequently, we expect
$m\propto \sqrt{t}$ to ensue after some time since
$m\propto -\int dh f(h)$ and this integral grows as $\delta w$ after 
$f$ nearly vanishes within the TW [see Fig. \ref{ftres}(b)], and
remains nearly constant therein subsequently.

The $m\propto \sqrt{t}$ growth stage comes to an end when $\delta w$ 
becomes as wide as the field distribution over all spins. Then
$m$ levels off to $m=m_s$. If however the magnetic system can 
exchange energy with the lattice at very low temperature, then a
thermally driven magnetization process eventually takes over. This is 
the subject of Sec. \ref{heat}

\subsection{Annealing}
\label{anneal}
  Assume that either $T_a\gg T_0$ or that the time spent in
the waiting stage is so short that the PDF that a randomly chosen 
site have field $h$ is not drastically different
from the PDF, $p(h)$, for a totally random spin 
configuration.\cite{gauss} On the other hand, the conditional 
probability to find
$\pm S$ on a site where the field is $h$ fulfills, in equilibrium, 
$p(\pm S\mid h)\propto \exp (\pm h/k_BT)$. Now, since the
joint distribution $p(\pm S,h)$ for finding, on a randomly chosen 
site, $h$ and $\pm S$ is in general given by
$p(\pm S,h)=p(\pm S\mid h)p(h)$,
\begin{equation}
p(\pm S,h)\propto p(h)e^{\pm h/k_BT}
\label{x2}
\end{equation}
follows in equilibrium. Monte Carlo results illustrate this point in 
Fig. \ref{fdos}(a). Therefore, since the mean energy is $\langle
E\rangle =(1/2)\int dh h [p_\downarrow (h)-p_\uparrow (h)]$, it follows that
$\langle E\rangle\simeq
\langle h^2 \rangle_0 /2k_BT$,\cite{xxx} where $\langle h^2 \rangle_0 
\equiv \int dh h^2 p(h)$. The replacement $
\langle h^2 \rangle_0/2k_BT\rightarrow\varepsilon_a$ generalizes the 
above equation to
\begin{equation}
p(\pm S,h)\propto p(h) e^{\pm 2h\varepsilon_a/\langle h^2 \rangle_0 }
\end{equation}
for all times up to equilibration. Then, to leading order in 
$\varepsilon_ah/\langle h^2 \rangle_0$,
\begin{equation}
p_\uparrow (h)-p_\downarrow (h)\simeq
2\frac{h\varepsilon_a}{\langle h^2 \rangle_0} p(h),
\label{otra}
\end{equation}
where $p_\uparrow (h)=p(S,h)$ and $p_\downarrow (h)=p(-S,h)$. 
Therefore, for Gaussian field distributions,\cite{gauss}
\begin{equation}
p_\uparrow (h)-p_\downarrow (h)\simeq
\sqrt{\frac{2}{\pi}}\frac{h\varepsilon_a}{ \sigma^3}e^{-h^2/2\sigma^2 },
\label{pmp}
\end{equation}
since $\langle h^2 \rangle_0 =\sigma^2$ then. All points for 
$[p_\uparrow (h)-p_\downarrow
(h)]/(h\varepsilon_a)$ obtained from MC calculations for SC lattices 
collapse onto a single curve in Fig. \ref{fdos}(b), in agreement
with Eq. (\ref{pmp}).

All results given below are for $\varepsilon_a/\sigma\lesssim 0.15$. 
This can be accomplished by annealing at $T\gtrsim 4T_0$. In
addition, only applied fields much smaller than
$\sigma$ are used. Thus, significant higher--order (in 
$\varepsilon_ah/\sigma^2$) contributions are avoided.

Note that Eqs. (\ref{x2})--(\ref{pmp}) are valid only in the 
annealing stage. They are inapplicable after quenching, because
spins are then not free to adjust to local fields.

\subsection{Magnetization process}
\label{mproc}
We now examine the system's time evolution after cooling it abruptly, 
at time $t=0$, to a temperature below, roughly,
$ 0.1U/(Sk_B)$. A field $H$ is applied for all $t>0$.
Then, real spin flips up to $\mid S_z\mid <\,S$ states can be neglected, and
tunneling through the ground state doublet is the only available path 
for spin reversals. Accordingly, spin flips are allowed only if
the spin's Zeeman energy is within the tunnel window. Now, if either 
the system is in thermal
contact with a reservoir at temperature $T$ such that 
$k_BT\gg\varepsilon_w$, or the energy is constant and sufficiently
high such that $k_BT_u\gg\varepsilon_w$, then
\begin{equation}
\dot{m}(t)= 2\Gamma \int dh\, \eta (H+h)f(h;t),
\label{unoa}
\end{equation}
where $f(h;t)\equiv p_\downarrow (h;t)-p_\uparrow (h;t)$. Let
\begin{equation}
\dot{\mu}(h,H;t)\equiv 2\Gamma \, \eta (H+h)f(h;t),
\label{uno}
\end{equation}
whose physical meaning follows from comparison of these two equations.

It is important to note that, Eq. (\ref{uno}) notwithstanding, 
$\dot{\mu}\neq -\dot{f}$. This is because $\dot{\mu}$ has to do with
numbers of spin flips, not with changes in dipolar fields, which are 
brought about by such spin
flips and also contribute to $\dot{f}$. In order to establish how $f$ 
evolves with time, it is convenient to define $g(h;t)\equiv
f(h,0)-f(h;t)$.  Now, $g(h;t)$ changes because spins which flip on 
sites where the field is within $h$ and $h+dh$ contribute
$\dot{\mu}(h,H;t)\Delta t dh$ to $gdh$ in time $\Delta t$, and
also because of dipolar field changes brought about by such flips. 
These two effects are approximately taken into account in,
\begin{equation}
g(h;t)\simeq\int_0^t d\tau\int d\tilde h G(h,\tilde h;t-\tau 
)\dot{\mu}(\tilde h ,H,\tau ),
\label{gh}
\end{equation}
where $G(h,\tilde h;t-\tau )$ is the PDF that, on a randomly chosen 
site, the field is $h$ at time $t$, given that the field on
that site was $\tilde h$ at time $\tau$. Again, note that 
$\dot{f}=-\dot{\mu}$ would only follow if field
redistributions, owing to spin flips, were disregarded, that is, if
$G(h,\tilde h;t-\tau )$ were replaced by $ \delta (h-\tilde h)$ in 
Eq. (\ref{gh}).

We next approximate $G(h,\tilde h;t-\tau )$. Recall that $H^2\ll 
\langle h^2\rangle_0$ throughout.
Consider a spin that points in opposite directions at times $t$ and $\tau$.
Let the fractional number of all such spins be $\phi (t,\tau)$. Since 
$p(h,t)$ is approximately constant for the times of interest
here, $\phi (t,\tau)\simeq \phi (t-\tau)$. For
$\phi (t-\tau )\ll 1$, we expect\cite{PWA}
\begin{equation}
G(h,\tilde h;t-\tau )\simeq \frac{u(t-\tau )}{\pi [(h-\tilde h 
)^2+u(t-\tau )^2]},
\label{greenH}
\end{equation}
where $u=2h_0\phi$, $h_0=(8\pi ^2/3^{5/2})g\mu_BS\rho_v$, $\rho _v$ 
is the number of spin sites per unit volume. Near the other end of
the time range, when
$\phi (t-\tau )\rightarrow 1/2$, we let\cite{gauss}
\begin{equation}
G(h,\tilde h;t-\tau )\simeq \frac{e^{-(h-\tilde h 
)^2/(2v^2)}}{\sqrt{2\pi }v(t-\tau )},
\label{green2}
\end{equation}
where $v=\sigma\sqrt{2\phi (t-\tau )}$, when $\phi (t-\tau )\rightarrow 1/2$.
\begin{figure}
\includegraphics[width=8cm]{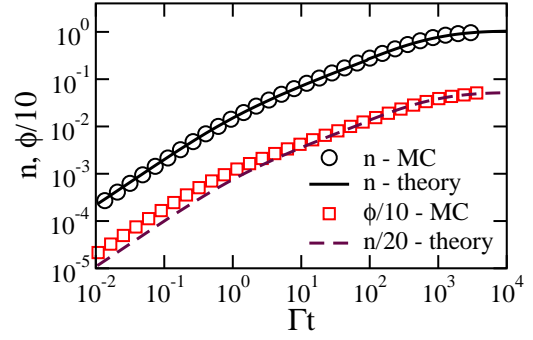}
\caption{Fraction of all spins $n$ that have flipped at least once, 
and quantity $\phi /10$ versus $\Gamma t$. In order to avoid
cluttering, $\phi /10$, instead of $\phi$ is plotted. All results are 
for $\varepsilon_w=0.1$. $\circ$ and $\Box$ stand for MC
averages over $5\times 10^4$ evolutions; the continuous and dashed 
lines follow numerically from Eq.
(\ref{mfinal2}). $n/20$ should equal $\phi /10$, according to Eq. 
(\ref{phin}).}
\label{nvst}
\end{figure}

To make progress, $\phi (t-\tau )$ must now be determined.
Let $\rho(h;t)$ be the PDF that a spin with a field $h$ acting on it 
has flipped at least once
in time $t$. In order to relate $\phi$ and $n$, given by
\begin{equation}
n(t)=\int dh\,\rho(h;t),
\label{n}
\end{equation}
we note that the total number of spin flips is much larger than the 
number of first time flips at all times. This implies that spins
flip several times before the dipolar field acting on them drifts 
away from the tunnel window. Accordingly, (see Fig. \ref{nvst}) we
adopt
\begin{equation}
\phi= n/2.
\label{phin}
\end{equation}
Finally, we write an
equation for
$\rho (h;t)$, from which
$n(t)$ and, consequently,
$\phi (t)$ then follow. Arguing as for  Eq. (\ref{gh}),
\begin{equation}
\rho(h;t)\simeq \int_0^t d\tau\int d\tilde h G(h-\tilde h;t-\tau 
)\dot{\nu}(\tilde h ,H,\tau ),
\label{rhoh}
\end{equation}
follows, where
\begin{equation}
\dot{\nu} (h ,H,\tau )\equiv [p(h,0 )-\rho(h,\tau )]\eta(h+H),
\label{nu}
\end{equation}
and we have assumed once again that $p(h;t)=p(h,0)$.

The desired equation for the magnetization process,
\begin{equation}
\frac{dx}{dt}\simeq c_1\frac{p(-H)}{p(0)}-c_2\int^t_0 d\tau
\frac{dx}{d\tau}\frac{2\varepsilon_w}{\omega (H;t-\tau)+2\varepsilon_w},
\label{mfinal2}
\end{equation}
where $x=m\sigma\langle h^2\rangle_0 /(\varepsilon_a\varepsilon_wH)$, 
$c_1=4\sqrt{2/\pi}$, $c_2=2$, and $\omega (H;t-\tau )\equiv
1/G(-H,-H;t-\tau)$, is derived in the Appendix. The above equation 
also holds for
$n(t)$ if we let $x=n\sigma/\varepsilon_w$, $c_1=\sqrt{2/\pi}$ and $c_2=1$.

For $H^2\ll \langle h^2\rangle_0$ and $H\ll h_0$,\cite{mT} $\omega 
(-H;t-\tau )\rightarrow u(t-\tau )$ in Eq. (\ref{greenH}) and
$\omega (-H;t-\tau )\rightarrow
\sqrt{2\pi}v(t-\tau )$ in Eq. (\ref{green2}). Then, 
\begin{equation}
\omega\simeq \min [2\pi h_0\phi, \sigma\sqrt{4\pi \phi} ]
\label{gfinal2}
\end{equation}
is an approximation that fits data obtained from simulations of field
distributions from various concentrations, $2\phi$, of randomly 
placed spins, in an otherwise empty lattice, reasonably well. In Eq.
(\ref{gfinal2}),
$\omega$ depends on
$t-\tau$ through
$\phi$ and, consequently, through $n$.

The functional dependence of $F$, shown in Eq. (2), follows from Eqs 
(\ref{mfinal2}) and (\ref{gfinal2}). To see how Eq. (3) comes
about, notice first that $\omega\ll \varepsilon_w$ when $\Gamma t\ll 
1$. Then, Eq. (\ref{mfinal2}) becomes $dx/d(\Gamma t)\simeq
c_1-c_2x$ for $H^2\ll \langle h^2\rangle_0$, whence Eq. (3) follows. 
For Eq. (4), we turn to {\it numerical} solutions of Eq.
(\ref{mfinal2}) which are exhibited in Figs. 1(a), (b), and (c).

We can obtain an analytic expression for $p$ from Eq. 
(\ref{mfinal2}). Since numerical solutions of Eqs. (\ref{mfinal2}) and
(\ref{gfinal2}) give
$x(t)\propto t^p$ for $\Gamma t \gtrsim 1$ and vanishingly small 
$\varepsilon_w/\sigma$, we let $\varepsilon_w/\sigma\rightarrow
0$ and try $m\propto t^p$ as a solution for all $t\gg \Gamma^{-1}$. 
Equations (\ref{phin}) and (\ref{gfinal2}) give
\begin{equation}
\omega\rightarrow \pi h_0n
\label{gfinal3}
\end{equation}
as $\varepsilon_w/\sigma\rightarrow 0$. Consider first $x=n\sigma 
/\epsilon_w$, and, consequently, $c_1=\sqrt{2/\pi}$ and $c_2=1$.
Then, for $H\ll\sigma$, Eq. (\ref{mfinal2}) becomes
\begin{equation}
\frac{dx}{dt}\simeq \sqrt{\frac{2}{\pi}}-\int^t_0 d\tau
\frac{dx(\tau )}{d\tau}\frac{1}{\alpha x(t-\tau )+1},
\label{mfinal5}
\end{equation}
where $\alpha =\pi h_0/(2\sigma )$. Assuming $x\sim t^p$, the change 
of variable $z=\tau /t$, brings Eq. (\ref{mfinal5}), in the
$t\rightarrow \infty$ limit, to
\begin{equation}
0= \sqrt{\frac{2}{\pi}}-\int^1_0 dz.
\frac{pz^{p-1}}{\alpha (1-z)^p},
\label{mfinal8}
\end{equation}
Equation (\ref{1}) follows immediately from integral tables 
\cite{grads}. We now return to Eq. (\ref{mfinal2}) to work with
$x=m\sigma\langle h^2\rangle_0 /(\varepsilon_a\varepsilon_wH)$,
$c_1=4\sqrt{2/\pi}$, and $c_2=2$. We now assume $x\sim 
t^{\tilde{p}}$, and, proceeding as above, obtain (1) $\tilde{p}=p$ 
and (2) $
m/n
\rightarrow
2\varepsilon_aH/\langle h^2\rangle_0$ as $t\rightarrow\infty$.
Therefore,
$p$ in
$m\propto t^p$ is given by Eq. (\ref{1}).

\begin{figure}
\includegraphics[width=8cm]{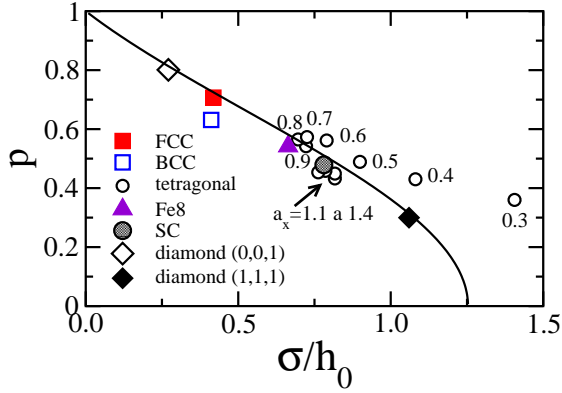}
\caption{Quantity $p$, defined by $m\propto t^p$, versus 
$\sigma/h_0$. Symbols stand for MC results for the shown lattices for
$\varepsilon_w=0.05$ and $H\lesssim \sigma /4$. The lattice structure 
for Fe$_8$ is as in Ref. \onlinecite{Fe_8}. The numbers next
to symbols for the tetragonal lattice stand for the ratio of the 
basal to the $z$--axis lattice constant. (0,0,1) and (1,1,1) stand
for easy magnetization axes. The full line stands for
Eq. (\ref{1}).}
\label{pend}
\end{figure}
Quantity $p$ from Eq. (\ref{1}), as well as results for $p$ obtained from MC
simulations for various lattices, are exhibited in Fig.
\ref{pend}. Monte Carlo generated data points are in reasonably good 
agreement with theory, that is, with
Eq. (\ref{1}), except for tetragonal lattices with $0.3$ and $0.4$ 
ratios of the basal to the $z$--axis lattice constant
(i.e., for $\sigma /h_0\simeq 1.08$ and $1.42$, respectively). This 
departure may follow from the fact that, for these rather
asymmetric lattices, $1/p(h=0)$ turns out to be nonlinear in the 
concentration $n$ of spin occupied lattice sites
for $ n\gtrsim 0.01$. This is in contradiction with Lorentzian field 
distributions\cite{PWA}
we have assumed for $n\ll 1$. Such distributions give $p(h=0)\propto 
n^{-1}$, which comes into Eq. (\ref{mfinal2}) through Eq.
(\ref{gfinal2}).

The relation $m/n\rightarrow 2\varepsilon_aH/\langle h^2\rangle_0$ 
holds just as well for small but finite
$\varepsilon_w/\sigma$, whence the constant $(1/2)\sqrt{\pi /2}$ in 
Eq. (5) follows, since $n\rightarrow 1$ as $t\rightarrow
\infty$. In addition, an intuitive argument that gives some insight into Eq. 
(5) is given in Sect. \ref{heat}.

\subsection{Tunnel window's imprint and field diffusion}
\label{TWI}
It is interesting to see how $g(h;t)$ behaves for early times, when 
$u\ll \varepsilon_w$ and $\phi \ll 1$. It then follows
from Eq. (\ref{greenH}) that $G(h)$ can be replaced by $\delta (h+H)$ 
in Eq. (\ref{gh}), which in turn yields
\begin{equation}
\dot{g}(h,H;t)=2\Gamma \eta (h+H)[f(h,0)-g(h;t)],
\label{B3}
\end{equation}
and, consequently, for $\Gamma t\ll 1$,
\begin{equation}
{g}(h;t)= f(h,0)\eta (h+H)(1-e^{-2\Gamma t}).
\label{B4}
\end{equation}
Therefore, when the variation of $f(h,0)$ over $\eta (h+H)$ is 
negligible, $g(h;t)$ is the tunnel window's
imprint. This is illustrated in Fig.
\ref{ftres}(b). A conjecture to this effect was made in Ref. 
\onlinecite{ww1}. In
order to show how long the condition
$u\ll
\varepsilon_w$, necessary for the validity of Eq. (\ref{B4}), holds, 
note that the total number of spins that flip in time $t$ is
approximately $p(h;0)\varepsilon_w\Gamma t$.  It follows immediately 
that $u\ll \varepsilon_w$ if $\Gamma t \ll
\sigma/h_0$. Therefore,  Eq. (\ref{B4}) holds while
$\Gamma t \ll 1$.\cite{hole3} This is illustrated in Fig. 
\ref{ftres}(b). Monte Carlo results for exponential TW shapes
that further illustrate this behavior are reported in Ref. \onlinecite{JMMM}.

After $\Gamma t\sim 1$, $g(h;t)$ no longer grows as in Eq. 
(\ref{B4}), with constant width. When $\Gamma t\gg 1$, the width of
$g(h;t)$ grows with time. A spin flip at any one site induces 
variations of dipolar fields at every other
site. Wherever a spin flipped at time $\tau$ the field $h$ evolves as 
in a random walk, away from $h=-H$ at time $\tau$, and
$G(h,-H;t-\tau )$ is its distribution at time $t$, as given in Eq. 
(\ref{greenH}) while $\phi(t-\tau ) \ll 1$. In this sense, one may
then speak of dipole {\it field diffusion}. It is the purpose of this 
section to calculate such field diffusion. A Fourier
transformation of Eq. (\ref{a2}) gives
\begin{equation}
g(k;t)\simeq \int_0^t d\tau G(k;t-\tau )\eta (bk)\frac{\dot{m}(\tau 
)}{2\varepsilon_w},
\label{x5}
\end{equation}
where $g(k;t)=\int dh \exp (-ikh) g(h;t)$, $G(k;t)=\int dh \exp 
[-ik(h-b\tilde h)] G(h,\tilde h;t)$, $\eta (bk)=\int dh \exp
[-ibk(h+H)]\eta (h+H)$, and $b=1-2\phi (t-\tau )$. A
similar equation [letting $g(k;t)\rightarrow \rho (k;t)$ and $m(\tau 
)\rightarrow n(t)$ in Eq. (\ref{x5})] obtains for $\rho (k;t)$.
Equation (\ref{x5}) can be solved numerically with the help of Eqs. 
(\ref{greenH})--(\ref{gfinal2}). Results obtained are shown in
Figs. \ref{ftres}(a) and (b). Perhaps it is worth pointing out that 
$p(h)$ changes appreciably only after very long
times when strong spin--spin correlations develop as the equilibrium 
ordered state is approached.\cite{Alonso,JMMM}

The spike in $f/h$, which develops at $h=0$ after some time, was not 
observed experimentally.\cite{ww1}
However, it is not a spurious effect. It is a consequence of the fact 
that the hole in $f$ spreads out in time, which
implies that $f(0,t)<0$ when $t>0$. The spike in $f/h$ at $h=0$ has a 
physical consequence: if $H$ is switched off at
some time $t\gtrsim 1$, a hole in $f(h,t)$ will develop at $h=0$, and 
the magnetization will decrease in time
accordingly.\cite{differ}
\begin{figure}
\includegraphics[width=8cm]{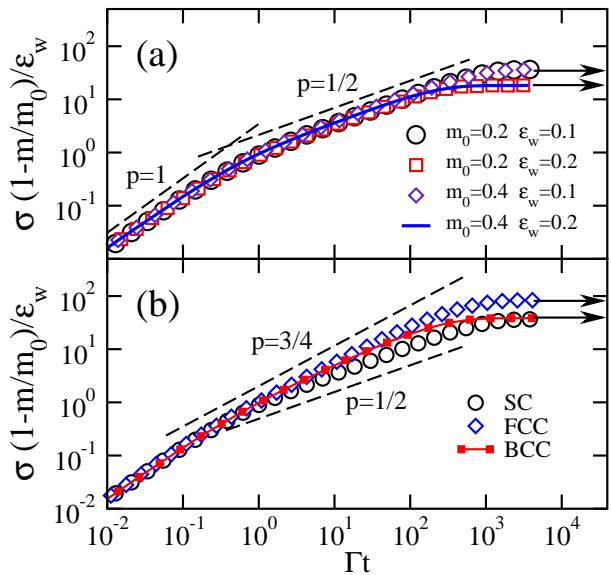}
\caption{(a) $\sigma (1-m/m_0)/\varepsilon_w$ versus $\Gamma t$ for 
zero field relaxation from weakly polarized states for
the shown values of $\varepsilon_w$ and of $m_0$.
Data points for $\Gamma t<64$ (for $64<\Gamma t<4\times 10^4$) follow 
from averages over $10^5$ (some $2\times 10^4$) MC simulations
of systems of $16\times 16\times 16$ ($8\times 8\times 8$) spins. (b) 
Same as in (a) but for BCC and FCC lattices,
$\varepsilon_w=0.1$ and $m_0=0.2$. The points indicated by the 
horizontal arrows correspond to $m\rightarrow 0$. The sloping
lines for
$t^p$, for the shown values of
$p$, are given for comparison purposes.}
\label{lastfb}
\end{figure}

\subsection{Relaxation in zero field}
\label{depol}
A slight variation of the experiment above is considered in this section. 
Suppose a magnetic field $H$ is applied during the
annealing stage, so that a polarization develops before quenching to 
very low temperatures, and let this applied field be switched
off at $t=0$, soon after quenching. Note that this differs from the 
problem considered in Refs. \onlinecite{PS} and
\onlinecite{vill} of systems that are initially {\it fully} 
polarized. Now, for a weakly polarized system, we expect
\begin{equation}
p_\uparrow (h;0)\simeq \frac{1}{2}(1+m_0)p (h;0),
\end{equation}
where $m_0$ is the initial magnetization, and similarly for 
$p_\downarrow (h;0)$ letting $m_0\rightarrow -m_0$.
For $H\ll 1$,
\begin{equation}
f(H,0)\simeq -\frac{m_0}{\sqrt{2\pi}\sigma},
\end{equation}
follows. Now, formally, replacement of
$\varepsilon_ah/\langle h^2\rangle_0$ by $-m_0/2$ in Eq. (\ref{otra}) 
gives the equation above. Since everything else works for the
relaxation of the magnetization the same as in Sec. \ref{mproc}, 
substitution of $\varepsilon_ah/\langle h^2\rangle_0$ by $-m_0/2$ in
Eq. (2) leads to
\begin{equation}
m\simeq m_0[1-b^\prime\varepsilon_w\sigma^{-1} F(\Gamma 
t,\sigma/\varepsilon_w,\sigma /h_0)],
\label{relax}
\end{equation}
where $b^\prime =2\sqrt{2/\pi}$ and $F$ is given by Eqs. (3)-(5). 
This is the desired equation.
Consequently, $m$ becomes vanishingly small after $\Gamma t\simeq
(\sigma /\varepsilon_w)^{1/p}$. As in sections above, $p\simeq 0.5$ for
SC lattices, but is different for other lattices. Up to then, $m$ 
relaxes, after $\Gamma t\approx 1$, as $t^p$. Monte Carlo results
illustrating this behavior are shown in Figs. \ref{lastfb}(a) and (b).

Thus, systems that are weakly polarized at quenching relax afterwards 
in complete analogy to
the magnetization process described in previous sections. Relaxation 
proceeds as $t^p$, from $\Gamma t\sim 1$ up to the time $(\sigma
/\varepsilon_w)^{1/p}$, when $m$ reaches a vanishingly small value. 
In addition, quantity $p$ depends on lattice structure the same as
above and as in Fig. \ref{pend}. This differs fundamentally from the 
relaxation of systems that are initially {\it fully} polarized,
which are the subject of Refs. \onlinecite{PS} and 
\onlinecite{exp,expb,vill}. In the latter condition: (1) the 
relaxation of the
magnetization depends crucially on system shape;\cite{C} (2) a 
$\sqrt{t}$ relaxation is predicted to be independent of lattice
structure; (3) this behavior holds while\cite{PS,exp} 
$\varepsilon_w/\sigma \lesssim \Gamma t$ and $m\gtrsim 0.9m_s$, which 
take
place much earlier than the magnetization processes that are the subject
of this paper.

\subsection{Late time magnetization}
\label{heat}

If the magnetic system does not exchange energy with the lattice, the 
magnetization finally levels off to the stationary value
$m_s\simeq 1.6 \varepsilon_aH/\langle h^2\rangle_0$, given by Eqs. 
(2) and (5). To gain some insight into this, note that: (1) $m=\int
dhg(h)$; (2) $g(h,t)$ comes near its stationary value, $f(-H,0)$, in 
the vicinity of $h=-H$ when $\Gamma t \gtrsim 1$;
afterwards, (3) $\int dhg(h,t)$ increases with time mostly because 
the width $\delta w$ of $g(h,t)$ increases up to the time when
$\delta w\approx 2\sqrt{\langle h^2\rangle_0}$, that is, when 
$g(h,t)$ becomes as wide as $p(h)$, and
reaches its stationary state. Then, $\int dhg(h)\approx 
f(-H,0)2\sqrt{\langle h^2\rangle_0}$, which, making use of Eq. 
(\ref{otra})
for
$f(-H,0)$ with
$\mid H\mid \ll 1$, and of $\sigma^2\sim \langle h^2\rangle_0$, gives 
the desired estimate for $m_s$. This comes to be when all spins
have flipped at least once, i.e., when
$n(t)\rightarrow 1$. The time evolution of $n(t)$ is shown in Fig. 
\ref{nvst} for $\varepsilon_w=0.1$.

\begin{figure}
\includegraphics[width=8cm]{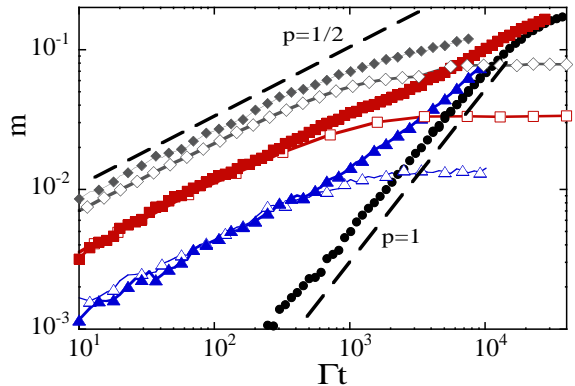}
\caption{$m$ versus $\Gamma t$ for $\varepsilon_w=0.1$, and $H=1$. 
All data points are for MC
simulations of $16\times 16\times 16$ spin systems. Open (full) 
symbols stand for evolutions in which heat exchange between magnetic
and lattice systems does not (does) take place. $\bullet$ is for 
$\varepsilon_a=0$, $\triangle$ is for $\varepsilon_a=0.10$,
$\blacktriangle$ is for $\varepsilon_a=0.097$, $\Box$ is for 
$\varepsilon_a=0.29$,
$\blacksquare$ is for $\varepsilon_a=0.29$, $\Diamond$ is for 
$\varepsilon_a=0.58$, and $\blacklozenge$ is for $\varepsilon_a=0.58$.
Data points stand for averages over at least $10^3$ MC runs. For 
comparison purposes, two straight dashed lines for $m\propto
t^p$ are shown.}
\label{mth}
\end{figure}

On the other hand, even when $k_BT\lesssim 0.1U/S$, heat exchange 
between magnetic and lattice systems does take place in some
systems of SMM's.\cite{metes} Then, the magnetization can increase 
further with time as the system evolves towards
its thermal equilibrium state. As has previously been 
shown,\cite{Ising} a relaxation time $\tau_{th}$ can then be defined, 
which is
given by
\begin{equation}
\Gamma\tau_{th} \approx \left(\frac{\sigma }{\varepsilon_w}\right) ^3.
\end{equation}
Accordingly, we expect
\begin{equation}
m(t) \simeq c\frac{H}{\sigma}\frac{t}{\tau_{th}}
\label{lt}
\end{equation}
to hold when thermalization effects become dominant. Equation (\ref{lt}), with
$c\simeq 1.5$, gives a rough fit of the nearly linear in time pieces 
in Fig. (\ref{mth}). A crossover to this late time regime, which
is driven by energy exchange with the lattice is easily appreciated 
in Fig. (\ref{mth}). The corresponding crossover time $\tau_{co}$
is best expressed in terms of another crossover time, $\tau_s$, when 
$m\propto\sqrt {t}$ crosses over to $m=m_s$. $\tau_{co}$ follows
from Eqs. (1)-(5) and Eq. (\ref{lt}). For 
$\varepsilon_a\gg\varepsilon_w$, the thermally driven linear behavior 
appears after $m$
levels off at $m_s$. Letting $\sigma^2\sim \langle h^2\rangle_0$ and 
$p\approx 2$,
\begin{equation}
\frac{\tau_{co}}{\tau_s}\sim \frac{\varepsilon_a}{\varepsilon_w}.
\label{tco1}
\end{equation}
obtains. On the other hand, if
$\varepsilon_a\ll\varepsilon_w$, linear behavior appears early, and 
magnetization leveling off is preempted. By the same
approximations,
\begin{equation}
\frac{\tau_{co}}{\tau_s}\sim\left(\frac{\varepsilon_a}{\varepsilon_w}\right)^2,
\label{tco2}
\end{equation}
then follows.

Even later yet, when $t\gtrsim \tau_{th}$, the magnetization crosses 
over to another regime as long--range order sets in. This can be
appreciated in Fig. \ref{mth} for $\Gamma t\gtrsim 10^4$.

\section{conclusions}
\label{dis}

We have given MC and theoretical evidence to show that the $m\propto 
\sqrt{t}$ behavior
observed in experiments on Fe$_8$ crystals \cite{ww1} after quenching 
and applying a small field $H$ at $t=0$ is driven by
correlations, which are previously established in the system while 
cooling to very low temperatures. Of the cooling protocol, only the
correlation energy $-\varepsilon_a$ at the time of quenching matters. 
The whole magnetization
process is ruled by Eqs. (2)-(5) when the magnetic system is 
thermally isolated.

The $m\propto \sqrt{t}$ behavior has been shown to be nonuniversal. 
Both MC and theory show that this behavior
ensues in SC lattices. Our MC simulations of an Fe$_8$ model, in the 
appropriate lattice,\cite{erratum} give $m\propto
\sqrt{t}$, in agreement with experiment.\cite{ww1} However, MC 
simulations of FCC and BCC lattice systems give
$m\propto t^p$, where $p$ varies with lattice structure as 
illustrated in Fig. \ref{funo}(c).
Numerical solutions of Eq. (\ref{mfinal2}) show that $p$ depends on 
lattice structure through $\sigma/h_0$ ($h_0$, which is defined
below Eq. (5), is proportional to the number of spin sites per unit 
volume) as shown in Fig.
\ref{pend}. Values of $\sigma$ and $p(0)$ are given in Table I for a 
few lattices. The $m\propto t^p$ regime has been shown
to cover the time range $1\lesssim \Gamma t\lesssim 
(\sigma/\varepsilon_w)^{1/p}$.

Similar results, namely Eq. (\ref{relax}), have been obtained for the 
magnetization relaxation in zero field of systems that have been
previously cooled in {\it weak} fields. This differs from the 
relaxation of {\it fully} polarized initial states, where the
$1-m\propto
\sqrt{t}$ behavior has been shown to be universal for $1-m\lesssim 
0.1$.\cite{PS} More remarkably, our results differ qualitatively
from the exponential relaxation predicted in Ref. \onlinecite{PS} for 
{\it weak} field cooled systems.

In order to test for system shape, size, and boundary effects, we 
have also performed MC simulations on spherical and box
like systems of various sizes, with periodic and with free boundary 
conditions on SC lattices. Quantity $p$ for
box like systems with PBC appears to be size and shape independent as 
long as there are more than four spins on each side. Box like
systems of
$L_x\times L_x\times L_z$ spins with free boundary conditions give 
values of $p$ that appear to agree in the $L_x\rightarrow\infty$
limit if $L_z\geq 16$ as well as in the
$L_z\rightarrow\infty$ limit if $L_x\geq 8$. Results for spheres with 
free boundary conditions also go to the same macroscopic limit :
$p=0.49\pm0.01$ for SC lattices.

The magnetization process depends significantly on whether the 
magnetic system is thermally isolated or not only after some
crossover time $\tau_{co}$, given by Eqs. (\ref{tco1}) and 
(\ref{tco2}). This is illustrated in Fig. \ref{mth}. It might be
interesting to observe this effect on systems in which specific heat 
experiments have shown
that energy is exchanged with the lattice within reasonable times at 
low temperatures.\cite{metes}

The simple result of Sec. \ref{TWI} deserves a comment. It shows that 
experimentally observed holes in $f(h,t)$ can be used to
establish the value of $\Gamma$, since holes develop independently of 
$\varepsilon_w$ and of $\varepsilon_a$ while $\Gamma
t\lesssim 0.5$, i.e., while $f$ at the bottom of the hole is 
approximately larger than one half its value at the top. This is 
useful
because there is no other simple relation we know of that one can use 
in order to extract the value of $\Gamma$ from othe experimental
observations. For instance, application of Eq. (\ref{B4}) to  Fig. 4 
in Ref. \onlinecite{ww1} yields
$\Gamma^{-1}\simeq 24$ s for Fe$_{8}$.

Except for the $m\propto \sqrt{t}$ behavior, which has been observed 
in Fe$_8$ over a limited time range,\cite{ww1}, all the
predictions we make here have yet to be observed experimentally. 
Since time scales can be controlled, varying $\Gamma$, through the
application of a transverse field, experimental observation seems feasible.
\acknowledgments
We thank F. Luis and W. Wernsdorfer for stimulating remarks. Support 
from the Ministerio de Ciencia y Tecnolog\'{\i}a of Spain,
through grant No. BFM2000-0622, is gratefully acknowledged.

\appendix
\section{}
Equation (\ref{gh}) can be further simplified making use of the 
approximation $\dot{\mu}(\tilde h,H,\tau )\simeq \eta
(h+H)\dot{m}(\tau )/2\varepsilon_w$. Then,
\begin{equation}
g(h;t)\simeq \int_0^t d\tau \frac{\dot {m}(\tau 
)}{2\varepsilon_w}\int d\tilde h\,G(h,\tilde h;t-\tau )\eta (\tilde 
h+H)
\label{a2}
\end{equation}
A similar equation follows for $\rho (t)$, making the replacement 
${\nu}(\tilde h,H,\tau )\rightarrow \eta (h+H)n(\tau
)/2\varepsilon_w$ in Eq. (\ref{rhoh}).
Now, from Eq. (\ref{unoa}) and the definition of $g$,
\begin{equation}
\dot{m}\simeq 2\Gamma \int dh[f(h,0)-g(h;t)]\eta (h+H),
\label{a1}
\end{equation}
follows immediately, and, similarly for $\dot{n}(t)$ (letting 
$m\rightarrow n$, $f\rightarrow p$, and $g\rightarrow \rho$ above).
Substitution of Eq. (\ref{a2}), into Eq. (\ref{a1}), gives
\begin{equation}
\dot{m}\simeq 2\Gamma \left[2\varepsilon_w f(-H,0)-\int_0^t d\tau 
\frac{\dot {m}(\tau )}{2\varepsilon_w}W(t-\tau )\right],
\label{b1}
\end{equation}
where $\varepsilon_w\ll\sigma $ has been assumed, and
\begin{equation}
W\equiv \int dh\int d\tilde h \eta (h+H) G(h,\tilde h;t-\tau )\eta 
(\tilde h+H).
\end{equation}
Approximations on $W$ follow. First, $\Gamma (t-\tau )\ll 1 
\Longrightarrow u(t-\tau
)\ll\varepsilon_w$, since $u=2h_0\phi$, $n/2\leq \phi \leq n(1-n/2) 
$, and $n(t)\sim (\varepsilon_w/\sigma )\Gamma t$ if $\Gamma t\ll
1$, whence
$G(h,\tilde h;t-\tau )\rightarrow
\delta (h-\tilde h)$, and therefore
\begin{equation}
W\rightarrow 2\varepsilon_w
\label{rw1}
\end{equation}
then. By the same argument, $\Gamma (t-\tau )\gg 1 \Longrightarrow u(t-\tau
)\gg\varepsilon_w$, whence it follows that variations of $G$ over the 
tunnel window are then negligible (from the assumption that
$\varepsilon_w\ll \sigma$), and
\begin{equation}
W\rightarrow (2\varepsilon_w)^2 G(-H,-H;t-\tau )
\label{rw2}
\end{equation}
therefore. We now use
\begin{equation}
{W(t-\tau )}\simeq \frac{2\varepsilon_w} 
{[2\varepsilon_wG(-H,-H;t-\tau )]^{-1}+1}
\end{equation}
to interpolate between Eqs (\ref{rw1}) and (\ref{rw2}). Substituting 
this equation into Eq. (\ref{b1}) [and into the analogous
equation for $n(t)$] gives Eq. (\ref{mfinal2}), which is the desired equation.


\begin{thebibliography}{99}
\bibitem{Gauss0}For Gaussian distributions $\sigma ^2=\langle 
h^2\rangle_0$. Numbers we have obtained from MC simulations
for $\sigma$ and for
$\langle h^2\rangle_0$ are listed in Table I.
\bibitem{xxx} This is also the leading term in a high temperature, 
$T_u$, series expansion for the energy, $-\varepsilon_w$. Hence
the energy remains (on the average) equal to $-\varepsilon_w$ in a 
simulation at temperature $T_u$.
\bibitem{gauss} For some lattices, such as cubic ones, $p(h)$ is 
nearly gaussian. It is not so for the Fe$_8$ lattice, nor for
lattices in which sites along the easy magnetization axis are much 
more closely spaced than sites in other directions.
\bibitem{hole3} A ``hole'' in $p(h;t)$ that resembles $\eta (h)$ was 
shown in Ref. \onlinecite{Alonso} to develop at much later
times.
\bibitem{differ}Actually, the data plotted in Ref. 11 is not quite 
the same as $f/h$ [W. Wernsdorfer (private communication)],
which may explain why no hole was seen experimentally.
\end{thebibliography}
\end{document}